\documentclass[cits]{PoS}
\usepackage{amsmath}
\usepackage{amssymb}
\usepackage{bm}
\title{\vspace*{-6.5cm}
{\hfill \texttt{\footnotesize CERN-PH-TH/2013-270}}
\vfill
Complex $\varphi^4$ Theory at Finite Temperature and Density via Extended Mean Field Theory}

\ShortTitle{Complex $\varphi^4$ Theory at Finite Temperature and Density via Extended Mean Field Theory}

\author{\speaker{Oscar Akerlund}$^a$ and Philippe~de~Forcrand$^{ab}$\\
        \llap{$^a$} Institut f\"ur Theoretische Physik, ETH Z\"urich, CH-8093 Z\"urich, Switzerland\\
        \llap{$^b$} CERN, Physics Department, TH Unit, CH-1211 Geneva 23, Switzerland\\
        E-mail: \email{oscara@itp.phys.ethz.ch}, \email{forcrand@itp.phys.ethz.ch}}

\abstract{We review the Extended Mean Field (EMFT) approximation and apply it to complex, scalar $\varphi^4$ theory on the lattice. We determine the ($T,\mu$) phase diagram and study the critical properties of the Bose condensation transition, at zero and finite temperature.
In both cases we obtain results which agree very well with recent Monte Carlo studies.
Within our approximation we can reach the thermodynamic limit and can thus obtain results at lattice spacings unreachable by present Monte Carlo simulations. We find that our approximation reproduces accurately many phenomena of the model, like the ``Silver Blaze'' behavior at zero temperature and dimensional reduction at finite temperature.}

\FullConference{31st International Symposium on Lattice Field Theory LATTICE 2013\\
		 July 29 - August 3, 2013\\
		 Mainz, Germany}

\newcommand{\rd}{\ensuremath{\mathrm{d}}}

\providecommand{\abs}[1]{\lvert#1\rvert}

\newcommand{\expv}[1]{\left\langle#1\right\rangle}

\newcommand{\bphi}{\bm{\Phi}}
\newcommand{\bdphi}{\bm{\delta\Phi}}
\newcommand{\be}{\begin{equation}}
\newcommand{\ee}{\end{equation}}
\newcommand{\benn}{\nonumber\begin{equation}}
\newcommand{\eenn}{\nonumber\end{equation}}
\def\bea{\begin{eqnarray}} \def\eea{\end{eqnarray}}
\def\beann{\begin{eqnarray*}} \def\eeann{\end{eqnarray*}}
\def\lsim{\raise0.3ex\hbox{$<$\kern-0.75em\raise-1.1ex\hbox{$\sim$}}}
\def\gsim{\raise0.3ex\hbox{$>$\kern-0.75em\raise-1.1ex\hbox{$\sim$}}}

\newcommand*\xbar[1]{%
  \hbox{%
    \vbox{%
      \hrule height 0.5pt 
      \kern0.3ex
      \hbox{%
        \kern-0.1em
        \ensuremath{#1}%
        \kern-0.1em
      }%
    }%
  }%
}

\begin{document}

\section{Introduction}\label{sec:introduction}\noindent
One of the most daunting problems in lattice field theory and computational physics is the so called ``sign problem'' which severely restricts the application of otherwise powerful Monte Carlo algorithms.
A ``sign (or phase) problem'' might occur for two reasons. Firstly, the statistics of the appearing fields might cause some configurations to carry a negative (fermions) or complex (anyons) weight. In principle it is possible to consider suitable subsets of the configuration space to end up with only non-negative weights \cite{Bloch:2013} but finding the appropriate subsets may be highly non-trivial.
Secondly, the action itself can be complex leading to sign problems arising also in bosonic systems. This may happen for example when a chemical potential is introduced.
The chemical potential couples to a conserved charge and breaks particle/anti-particle symmetry, which renders the action complex.
In order to have a conserved charge the Lagrangian needs to have a continuous global symmetry, the simplest being a $U(1)$ symmetry. One simple and well-studied $U(1)$-symmetric model is the complex $\varphi^4$-theory, which models a relativistic Bose gas. Complex $\varphi^4$ is also related to the Standard Model Higgs boson which is a two-component complex scalar with a quartic self-interaction.

One promising method for dealing with the sign problem is Complex Langevin (CL)~\cite{Aarts:2009yj}. Furthermore, the sign problem can sometimes be avoided by a variable transformation (cf. world-line Monte Carlo \cite{Endres:2006xu, Gattringer:2012df}) but a suitable set of variables is not always easy to find. Mean Field methods, although approximate, can most of the time use the symmetries of the Lagrangian to rotate the fields and make the action real. Above the upper critical dimension mean field theories typically yield fair predictions for local variables but usually over-emphasize the ordered phase.
Another shortcoming of mean field theory is that it can not be used to determine correlation functions and can not be used to study finite temperature.
EMFT \cite{Pankov:2002, Akerlund:2013} is an extension of mean field theory which incorporates a higher degree of self-consistency and can be used to overcome these limitations.

In this proceeding we apply EMFT to the $U(1)$-symmetric $\varphi^4$-theory. In particular we study the model at nonzero temperature and density, the main result being the $(T,\mu)$-phase diagram. As much as possible we compare our results with recent Monte Carlo~\cite{Gattringer:2012df} and complex Langevin~\cite{Aarts:2009yj} results.

\section{Lattice Action and Extended Mean Field Theory (EMFT)}\label{sec:emft}\noindent
EMFT \cite{Akerlund:2013} is based on the work of Pankov \cite{Pankov:2002}. It is an extended version of mean field which takes also some fluctuations into account. More precisely it resums all local diagrams contributing to the free energy \cite{Pankov:2002}. We have previously applied this approximation to the real scalar $\varphi^4$-theory with very good results \cite{Akerlund:2013}.
This extended version of mean field is equivalent to the local limit of Dynamical Mean Field Theory (DMFT) which is extensively used in the condensed matter community, see eg. \cite{Georges:1996} for a review. In DMFT the effective Weiss field is a function of one coordinate, usually [Euclidean] ``time'' (hence the name), and emulates effective, nonlocal interactions. The effective model is a world line frozen in space and the effective field is determined self-consistently by matching the $1d$ Green's function to an approximation of the Green's function of the full theory.
In the ``local time'' limit the world line is just one point and the effective field is a number. However, this local limit is different from usual mean field since it contains an effective field which couples to $\varphi^2$. Aarts~\cite{Aarts:2009hn} has also presented a mean field version of the complex Langevin equations which involves the two-point correlator. His method explicitly takes a tadpole correction proportional to $\lambda$ into account and works well when $\lambda$ is small.
EMFT, on the other hand, resums local diagrams to all orders in $\lambda$ and works well for all values of the coupling \cite{Akerlund:2013}.

We proceed by deriving the effective action and the self-consistency equations. We will write the well-known lattice action of complex $\varphi^4$-theory with chemical potential $\mu$ in $d$ dimensions using a slightly unconventional notation more convenient for our purposes,
\be
S = \displaystyle\sum_x\left[\!-\sum_\nu \bphi^\dagger_{x +\widehat{\nu}}\bm{E}(\mu\delta_{\nu,t})\bphi_x \!+\! \frac{\eta}{2}\abs{\bphi_x}^2 \!+\! \frac{\lambda}{4}\abs{\bphi_x}^4\!\right], 
\ee
with 
\begin{equation}
\bphi^\dagger = (\varphi^*,\varphi),\;\; \bm{E}(x) = \begin{pmatrix} e^{-x} & 0 \\ 0 & e^{x} \end{pmatrix},\;\;\eta = m_0+2d.
\end{equation}
We see that in the free case ($\lambda=0$) the action is quadratic in $\bphi$ and the connected Green's function in Fourier space is readily obtained,
\begin{equation}
\bm{G}_0(k) = \expv{\bphi\bphi^\dagger}_c =  \begin{pmatrix} \expv{\varphi\varphi^*} & \expv{\varphi\varphi} \\ \expv{\varphi^*\varphi^*} & \expv{\varphi^*\varphi} \end{pmatrix}  - \langle\varphi\rangle^2 = \left(\eta - 2\sum_{\nu=1}^d\cos\left(k_\nu - i\mu\delta_{\nu,t}\right)\right)^{-1}\bm{I}_{2\times2}.
\end{equation}
This is a central point of EMFT and similar methods. We know the Green's function at some point in parameter
space, here, $\lambda=0$, and we can quantify the deviation of the full Green's function from the known one in the form of a function that only depends on the
interaction, in this case $\lambda$. The goal is then to find some simpler, but in some sense equivalent, model which 
we can solve completely and which shares the same interaction-dependent deviation. In most cases we can only find a simple model
with an approximately equal deviation, or one which is only valid in some limiting regime. 
In this spirit, we express the full lattice Green's function as,
\be\label{eq:gf_latt}
\bm{G}^{-1}(k) = \bm{G}_0^{-1}(k) - \bm{\Sigma}(k),
\ee
where $\bm{\Sigma}$ is the self-energy due to $\lambda$. We will now as usual expand $\bphi$ around its (real) mean, $\expv{\bphi}=\xbar{\bm{\phi}}$: $\bphi = \xbar{\bm{\phi}} + \bdphi$. Concentrating on the field at the origin, $\bphi_0$, we can write the action as,
\begin{align}
S &= \frac{\eta}{2}\abs{\bphi_0}^2 + \frac{\lambda}{4}\abs{\bphi_0}^4 - 2\xbar{\bm{\phi}}^T\bphi_0(d-1+\cosh(\mu))\nonumber\\
&\phantom{=}-\sum_\nu\left(\bdphi_{0+\hat{\nu}}^\dagger\bm{E}(\mu\delta_{\nu,t})\bdphi_0+\bdphi_{0}^\dagger\bm{E}(\mu\delta_{\nu,t})\bdphi_{0-\hat{\nu}}\right) + S_\text{ext} = S_0+\delta S+S_\text{ext},\label{eq:action_onesite}
\end{align}
where the last term $S_\text{ext}$ does not depend on $\bphi_0$ and is irrelevant to our purpose. The middle term $\delta S$ describes the interaction of the field $\bphi_0$ at the origin, which we keep fully dynamical (``live''), with the field at neighboring sites, $\bphi_{0\pm\hat{\nu}}$, which we collectively denote by $\varphi_\text{ext}$ and want to integrate out. This is realized by replacing $\delta S$ by its cumulant expansion
\begin{equation}\label{eq:zext}
Z = \int\rd\varphi_0\mathcal{D}\varphi_\text{ext}\,e^{-S_0(\varphi_0)-\delta S(\varphi_0,\varphi_\text{ext})-S_\text{ext}(\varphi_\text{ext})}=\int\rd\varphi_0\,e^{-S_0(\varphi_0)-\expv{\delta S}_\text{C,ext}(\varphi_0)},
\end{equation}
under the action $S_\text{ext}$. Up to now everything is exact but in practice we will have to truncate the cumulant expansion at some order. To second order in the fluctuation, $\bdphi_0$, it reads:
\begin{align}
\expv{\delta S}_\text{C,ext}\approx&\expv{\sum_{\pm\nu}\bdphi_{\hat{\nu}}^\dagger\bm{E}(\pm\mu\delta_{\nu,t})\bdphi_0}_{\!\!\!S_\text{ext}} \nonumber\\
&+ \frac{1}{2}\expv{\sum_{\pm\nu}\bdphi_{\hat{\nu}}^\dagger\bm{E}(\pm\mu\delta_{\nu,t})\bdphi_0 \sum_{\pm\rho}\bdphi_{\hat{\rho}}^\dagger\bm{E}(\pm\mu\delta_{\rho,t})\bdphi_0}_{\!\!\!S_\text{ext}} \!\!\!\!\!= 0 + \frac{1}{2}\bdphi_0^\dagger\bm{\Delta}\bdphi_0\label{eq:cum_exp}
\end{align}
The first term is zero by construction and $\bm{\Delta}$ is a $2\times2$ unknown matrix which represents the second term and will be determined self-consistently. Since $\bphi$ has only two degrees of freedom ($\text{Re}\varphi$ and $\text{Im}\varphi$) so has $\bm{\Delta}$ which is why we can choose it to be real, symmetric and with $\bm{\Delta}_{11}=\bm{\Delta}_{22}$. We restrict ourselves to second order in $\bdphi$ for simplicity. In principle, expanding to higher order provides a way to systematically improve the approximation. This second order term has a clear interpretation; it represents all loops propagating in the external bath and closing at the origin. Making the above substitution in Eq.~\eqref{eq:action_onesite} and using $\bdphi_0 = \bphi_0+\xbar{\bm{\phi}}$ yields the effective, one-site action:
\begin{equation}\label{eq:action_emft}
S_\text{EMFT} = \frac{1}{2}\bphi_0^\dagger\left(\eta\bm{I} - \bm{\Delta}\right)\bphi_0 + \frac{\lambda}{4}\abs{\bphi_0}^4 - 2\xbar{\phi}\text{Re}[\varphi](2(d-1+\cosh(\mu))-\bm{\Delta}_{11}-\bm{\Delta}_{12}).
\end{equation}

The EMFT Green's function can, like the full Green's function $\expv{\bdphi_0\bdphi_0^\dagger}$, be defined as a free part and a self-energy,
\be
\bm{G}^{-1}_\text{EMFT} = \eta\bm{I} - \bm{\Delta} - \bm{\Sigma}_\text{EMFT}.
\ee
The mapping is complete by neglecting the $k$-dependence of the full self-energy in Eq.~\eqref{eq:gf_latt} and replacing it with the local, i.e. $k$-integrated self-energy which is approximated by the EMFT self-energy $\bm{\Sigma}_\text{EMFT}$. This is justified since if we had taken the entire cumulant expansion in Eq.~\eqref{eq:cum_exp} then the effective action would exactly correspond to the full theory and would generate all point-like observables. Substituting $\bm{\Sigma}_\text{EMFT}$ into Eq.~\eqref{eq:gf_latt} yields,
\be\label{eq:gf_sub}
\bm{G}^{-1}(k) \approx \bm{G}^{-1}_\text{EMFT} + \bm{\Delta} -2\sum_{\nu=1}^d\cos\left(k_\nu - i\mu\delta_{\nu,t}\right)\bm{I}.
\ee
$\bm{\Delta}$ is self-consistently determined by demanding that the (approximate) local full lattice Green's function $\bm{G}_{xx} = \int\frac{\rd^dk}{(2\pi)^d}\bm{G}(k)$ equals the EMFT Green's function. This is the same as finding the stationary point of the (approximate) local free energy functional~\cite{Potthoff:2003}. This, together with the self-consistency of $\xbar{\phi}$ yields a set of three coupled self-consistency equations,
\begin{align}
\xbar{\phi} &= \expv{\varphi}_{S_\text{EMFT}},\\
\int\frac{\rd^dk}{(2\pi)^d}\bm{G}(k) &= \bm{G}_\text{EMFT},\label{eq:sc_gf}
\end{align}
where the matrix equation Eq.~\eqref{eq:sc_gf} contains two independent equations, one for the diagonal element and one for the off-diagonal. The momentum integral turns into a finite sum in the case of finite volume or temperature. Note that, as an output, one obtains an approximation to the Green's function via Eq.~\eqref{eq:gf_sub}, in addition to $\expv{\varphi}$ and $\bm{\Delta}$. As usual when dealing with self-consistently determined parameters we solve the above equations iteratively. There is, however, some freedom in the choice of numerical procedure and some are more efficient than others. For more details, see \cite{Akerlund:2013}.

\section{Results}
\begin{figure}[t]
\centering
\includegraphics[width=0.45\linewidth]{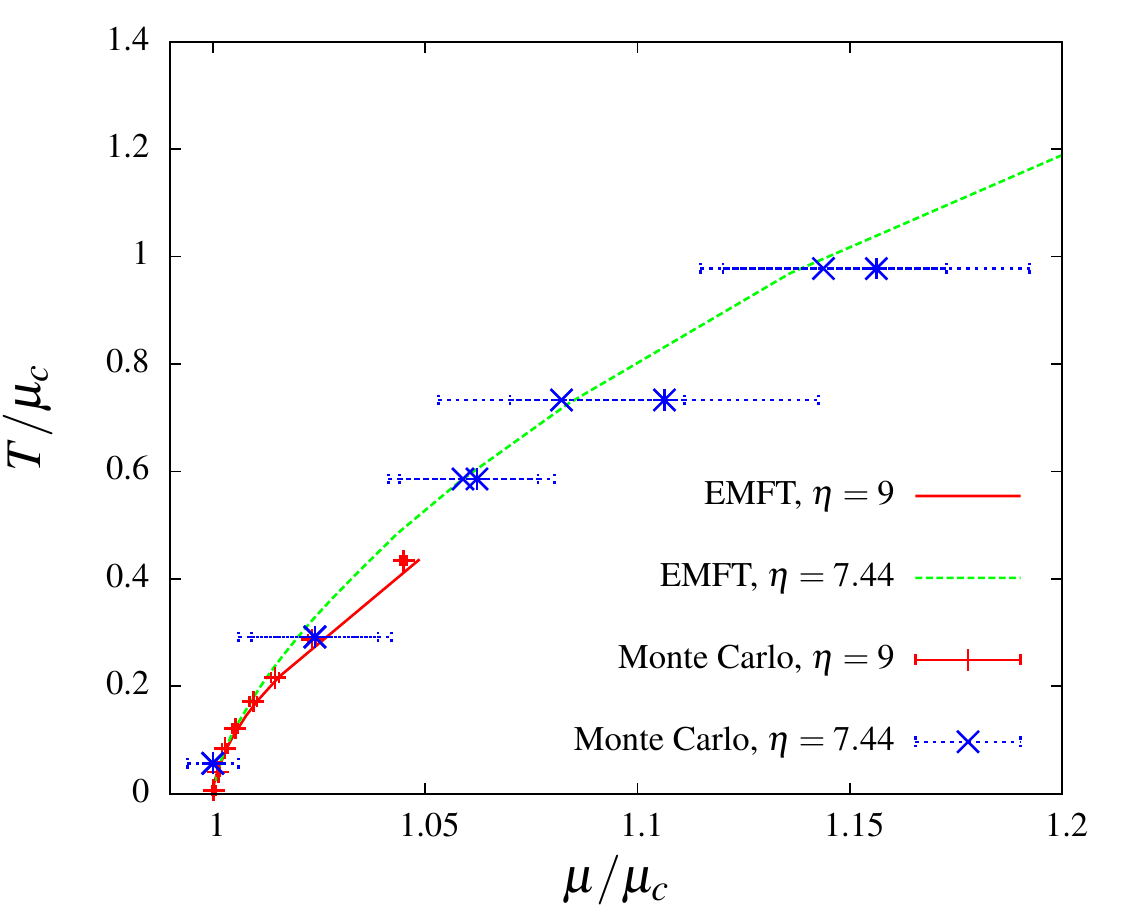}
\caption{$(T/\mu_c(T=0),\mu_c(T)/\mu_C(T=0))$ phase diagram of complex $\varphi^4$-theory at $\lambda=1$ obtained by EMFT and by world-line Monte Carlo \cite{Gattringer:2012df}. We have used the same two values of $\eta$ as in \cite{Gattringer:2012df} and the results agree very well for both.}
  \label{fig:phasediag}
\end{figure}

\noindent
One very interesting advantage of EMFT over standard mean field theory is the possibility to study finite temperature effects on the model in question. This is simply achieved by truncating the sum over $k_t$ in Eq.~\eqref{eq:sc_gf} at some finite value of $N_t$. This allows us to define a temperature in lattice units, $aT=N_t^{-1}$, or in terms of the chemical potential, $T/\mu=((a\mu)N_t)^{-1}$. We can then solve the self-consistency equations and obtain all observables as a function of the temperature. The most important result is perhaps the $(T/\mu_c,\mu/\mu_c)$ phase diagram which we show in Fig.~\ref{fig:phasediag}. $\mu_c$ is defined to be the critical chemical potential at zero temperature. We show the phase diagram for $\eta=9$ and $\eta=7.44$ to allow for a direct comparison with Monte Carlo results obtained by Gattringer and Kloiber \cite{Gattringer:2012df}. These authors used a world-line formulation of the partition function which has no sign problem, and used Monte Carlo to sample the configuration space. We see an excellent agreement at all temperatures and both values of $\eta$. We find $\mu_c(\eta=9)=1.1458$ and $\mu_c(\eta=7.44)=0.1720$ which should be compared with $\mu_c(\eta=9) = 1.146(1)$ and $\mu_c(\eta=7.44)=0.170(1)$ found in \cite{Gattringer:2012df}.

\begin{figure}[t]
\centering
\includegraphics[width=0.43\linewidth]{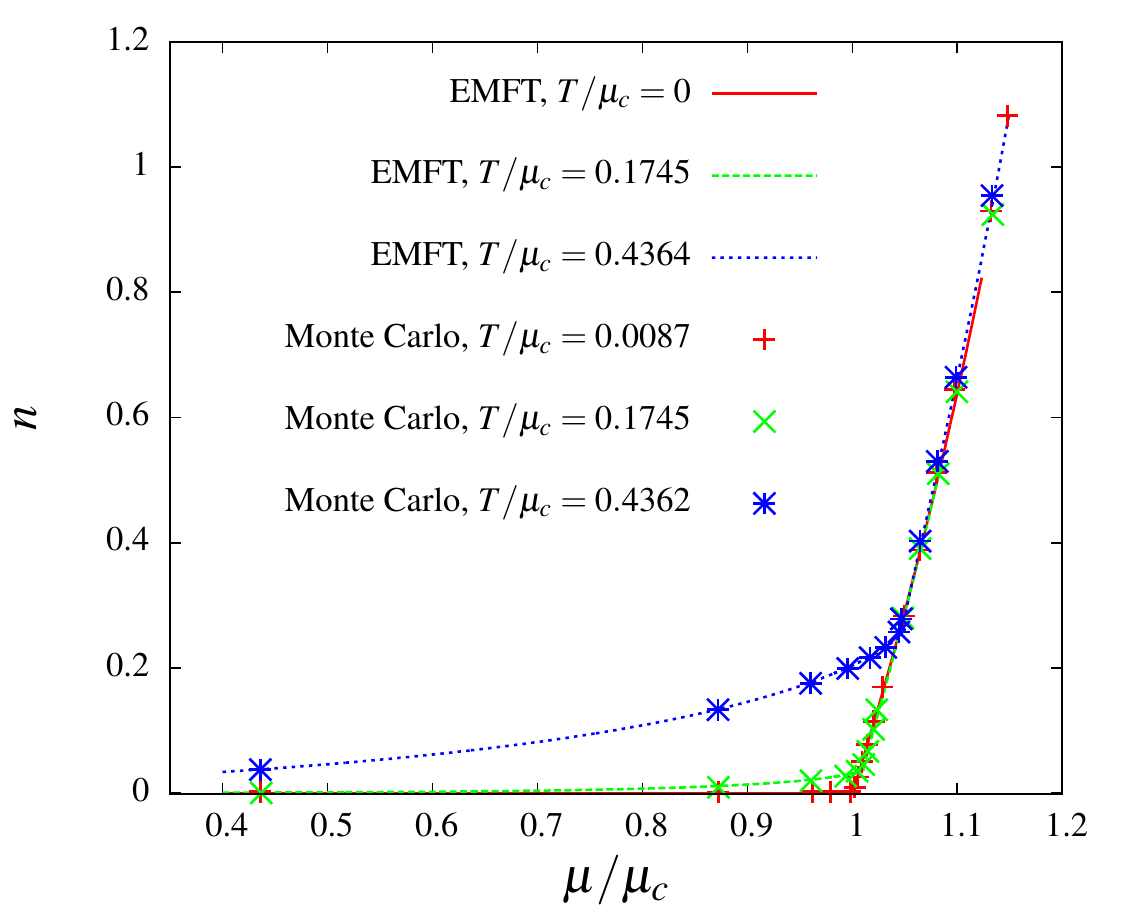}
\includegraphics[width=0.43\linewidth]{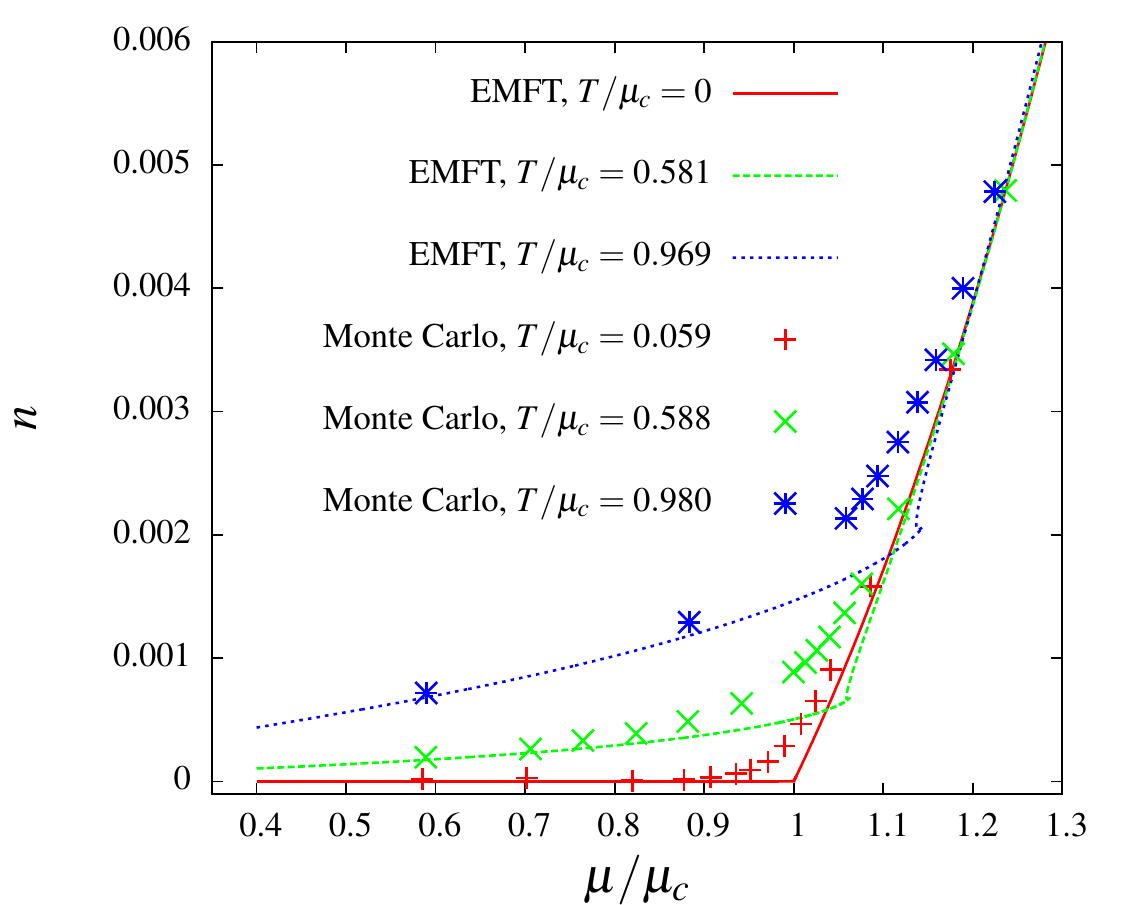}
\caption{The density $n$, as a function of $\mu$ for a few different temperatures at $\lambda=1$ and $\eta=9$ (\emph{left panel}) and $\eta=7.44$ (\emph{right panel}). We compare EMFT with Monte Carlo simulations \cite{Gattringer:2012df} on an $N_s^3\times N_t$ lattice, with $N_s=20$ for $\eta=9$ and $N_s=24$ for $\eta=7.44$. The temperatures correspond to $N_t=100(\infty),5,2$ and $N_t=100(\infty),10,6$ for $\eta=9$ and $\eta=7.44$ respectively. In EMFT, we take $N_t=\infty$ instead of $N_t=100$ for computational convenience. The EMFT results are obtained in the thermodynamic limit, i.e. $N_s=\infty$.}
  \label{fig:dens}
\end{figure}

Also quantitative comparisons of the density as a function of $\mu$ at different temperatures confirm the accuracy of EMFT. We compare again with the Monte Carlo simulations in \cite{Gattringer:2012df} with $\lambda=1$ and $\eta=9$ and $\eta=7.44$. The result is presented in Fig.~\ref{fig:dens}. At the larger value of $\eta$ (\emph{left panel}) the finite volume effects in the Monte Carlo data are small and the two methods agree almost perfectly with each other. Since the nonzero temperature contribution to the density is closely related to the Green's function at separation $a$, this shows that EMFT is not restricted to predicting the local Green's function $G_{xx}$. Closer to the continuum limit, at $\eta=7.44$ (\emph{right panel}), the finite-size effects in the Monte Carlo data are more severe, which manifests itself as a rounding of the phase transition. EMFT does not suffer from finite-size effects and shows a sharp transition. Away from the transition the two methods agree very well.

\subsection{First Order Transition}
\begin{figure}[t]
\centering
\includegraphics[width=0.43\linewidth]{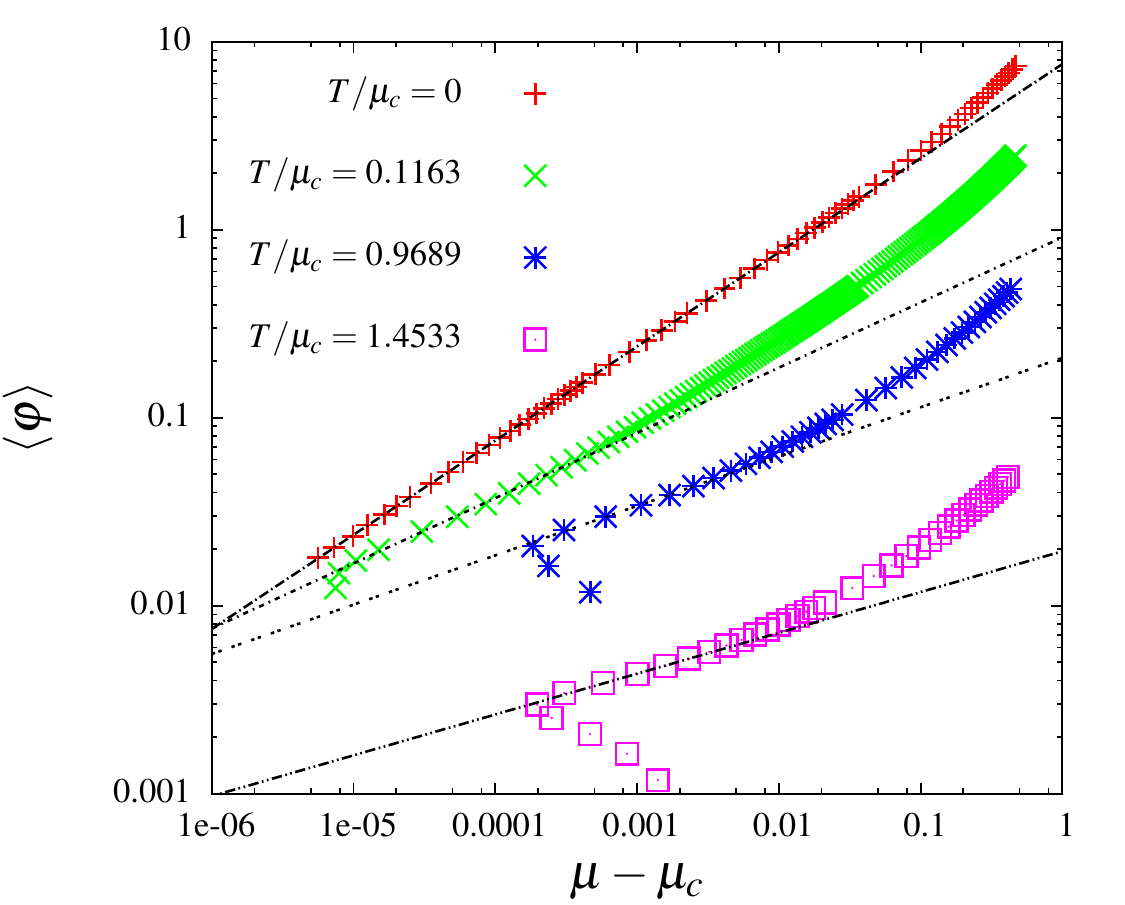}
\includegraphics[width=0.43\linewidth]{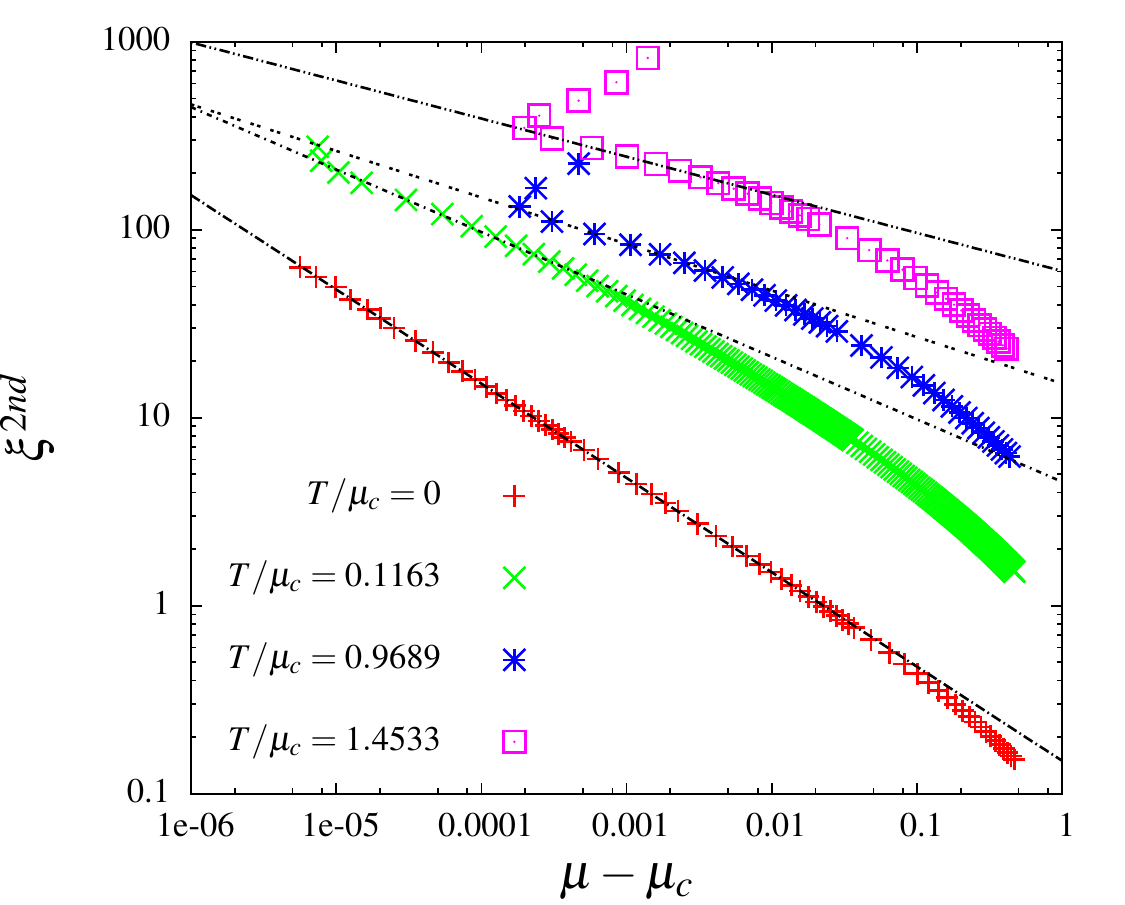}
\caption{The expectation value of the field (\emph{left panel}) and the correlation length (\emph{right panel}), as a function of $\mu$ for a few different temperatures at $\lambda=1$ and $\eta=7.44$. The curves are shifted vertically for readability. At zero temperature we find a second order transition with mean field exponents but, as soon as the temperature is turned on we find a weak first order transition.}
  \label{fig:first}
\end{figure}

If we let $\mu$ come very near its critical value, EMFT incorrectly predicts that the transition turns weakly first order as temperature is turned on. This can most easily be seen in Fig.~\ref{fig:first} where we show the expectation value $\langle\varphi\rangle$ of the field and the correlation length $\xi^{2nd}$ defined from the second moment method. We see both that a jump in $\langle\varphi\rangle$ develops and that the critical exponents change from mean field values to $1/d$. However, we also find that the free energy is discontinuous at the phase transition, so we conclude that the approximation fails and that the first order transition should be ascribed no physical meaning.
Although EMFT still produces quantitatively good predictions of various observables such as the critical chemical potential and the density, this is of course an undesired feature. In order to cure this behavior we would be required to go to higher orders in the cumulant expansion, Eq~\eqref{eq:cum_exp}.

\section{Conclusions}
We have seen that EMFT works excellently in four dimensions at zero temperature where it correctly predicts a second order phase transition with mean field exponents and a quantitatively very accurate value of the critical chemical potential.
EMFT also provides a computationally cheap method to probe the system at finite temperature, and although it incorrectly predicts a first order transition we obtain observables like the critical chemical potential and the density which numerically agree very well with state of the art Monte Carlo simulations \cite{Gattringer:2012df}.
This makes EMFT a potentially very useful tool for making predictions on the existence and location of a phase transition. However, EMFT might have no predictive power regarding the order of previously unknown phase transitions. Nevertheless, due to its simplicity and low computational cost, it can serve as a complement to more sophisticated methods.

A natural and straightforward next step could be to study a model containing a multi-component scalar field, for example a gauge-less $SU(2)$ Higgs model. An even more interesting extension would be to include the gauge field and study a $U(1)$ Higgs model. Since the smallest gauge independent object, the plaquette, lives on four lattice sites we would have to extend the model to work with a cluster of live sites, a direction which is interesting on its own since it allows for self-consistent determination of momentum-dependent variables.

\section{Acknowledgments}
This work is an extension of \cite{Akerlund:2013}, with Antoine Georges and Philipp Werner. We thank them both for valuable insights.

\end{document}